\documentstyle[preprint,aps]{revtex}
\draft
\begin{document}
\preprint{Phys. Rev. C (1996) in press.} 
\title{Kaon dispersion relation and flow in relativistic heavy-ion collisions}
\bigskip
\author{Bao-An Li and C. M. Ko }
\address{Cyclotron Institute and Physics Department,\\ 
Texas A\&M University, College Station, TX 77843}
\maketitle
 
\begin{quote}
Within the framework of a relativistic transport model (ART) for 
heavy-ion collisions at AGS energies, we examine the effects of kaon 
dispersion relation on the transverse flow of kaons and their 
transverse momentum and azimuthal angle distributions. We find that 
the transverse flow is the most sensitive observable for 
studying the kaon dispersion relation in dense medium. 
\end{quote}
\newpage

The properties of a kaon in the extremely hot and dense enviroment created 
in relativistic heavy-ion collisions have been subjects of many theoretical
studies(see, e.g.\ \cite{brown} for a recent review). Experimentally, 
to extract the in-medium kaon properties is one of the most challenging 
tasks at several heavy-ion facilities (e.g. \cite{grosse,js}). 
In a recent study it was first proposed that the kaon collective flow might be
a sensitive probe of the kaon dispersion relation in dense medium\cite{gqli1}.
Furthermore, within the RVUU model\cite{ko} it was shown that for 
heavy-ion collisions at SIS/GSI energies (E/A$\leq$ 2 GeV) the flow 
pattern of kaons is affected dramatically by the kaon potential in
medium\cite{gqli1,gqli2}. This suggestion has recently attracted much 
attention\cite{ritman}. In particular, at SIS/GSI significant efforts will be
devoted in the next few years to experimental studies of kaon flow 
and the extraction of the kaon dispersion relation in medium\cite{best}.  
It is an interesting question, as we were actually frequently asked at 
recent meetings, whether the effect also exists at higher energies.
In this short report, we extend the study to AGS energies
and show that at high energies the kaon flow is also a very sensitive probe of
the kaon dispersion relation. For comparison we also examine effects
of the kaon dispersion relation on its transverse momentum 
distribution and azimuthal angle distribution with respect to the
reaction plane (``squeeze-out''). Results of this study are expected 
to be useful for current experiments at Brookhaven's 
AGS (e.g. E859, E866 and E877 collaborations) and the long range plan of
GSI.

Our study is based on a relativistic transport model (ART) 
for heavy-ion collisions at AGS energies\cite{art}.
In this model the phase space distribution functions of
baryons ($N,\Delta(1232),N^{*}(1440),N^{*}(1535),\Lambda,\Sigma$)
and mesons ($\pi,\rho,\omega,\eta,K$) are evolved under the influence of 
hadron-hadron scatterings and also optional mean fields.
In transport models (see, e.g., \cite{li89,wang,bauer} for a review), 
the imaginary part of kaon self-energy is approximately treated by 
the scatterings of kaons with other hadrons, and the real part of 
the self-energy is given by the mean-field potential. Although 
various approaches have been used to evaluate 
the kaon dispersion relation in dense 
medium (e.g. \cite{brown,kn,bkr,pw,mt,bkk,lsw,ynm}, we shall use in the 
present study as an illustration the kaon dispersion relation 
determined from the kaon-nucleon scattering length $a_{KN}$ using the 
impulse approximation\cite{brown}, i.e.,
\begin{equation}
\omega(p,\rho_b)=[m_K^2+p^2-4\pi (1+\frac{m_K}{m_N})a_{KN}\rho_b]^{1/2},
\end{equation}
where $m_K$ and $m_N$ are the kaon and nucleon masses, respectively; $\rho_b$
is the baryon density and $a_{KN}\approx -0.255$ fm is the isospin-averaged 
kaon-nucleon scattering length.
The kaon potential in medium can then be defined as 
\begin{equation}
U(p,\rho_b)=\omega(p,\rho_b)-(m_K^2+p^2)^{1/2}.
\end{equation}

Effects of the above potential on experimental observables 
can be seen qualitatively from the force acting on a kaon during its
propogation
\begin{equation}
\vec{F}=\frac{d\vec{p}}{dt}=-\vec{\nabla}U
=\frac{2\pi}{\omega}(1+\frac{m_K}{m_N})a_{KN}\vec{\nabla}\rho_b.
\end{equation} 
Since the force is inversely proportional to the energy $\omega$, 
low energy particles are most likely to be affected by 
the kaon potential. This suggests that one should look for signatures 
of the kaon potential on kaon observables mainly around the midrapidity.
Moreover, the kaon potential tends to force kaons away from
the direction along which the baryon density increases. The latter 
reveals the underlying reason why the kaon transverse flow is a sensitive 
probe of its potential in medium. The transverse flow 
of baryons in heavy-ion collisions is an experimentally well 
established fact at energies even as high 
as 160 GeV/nucleon. Kaons gain some collectivity
not only from their interactions with baryons, but also because they are mainly
produced from baryon-baryon collisions in
heavy-ion collisions. The baryon transverse flow results in a large 
asymmetry of baryon density distribution in the reaction plane, 
especially in the later stage of the reaction. Kaons moving in this 
medium thus develope an opposite asymmetry due to their interactions 
with baryons through the mean-field potential. To demonstrate this 
effect we will compare in the following results obtained with and 
without the kaon potential.

It is now well-known that the best way to reveal transverse flow
in heavy-ion collisions is to perform the in-plane transverse momentum 
analysis\cite{dani}. We have recently performed this analysis for 
baryons and pions using the ART model for heavy-ion collisions at 
beam energies from SIS/GSI to AGS/Brookhoven energies\cite{art}. 
Here we concentrate on the analysis
for kaons. In Fig.\ 1 the average kaon transverse momentum (scaled by
kaon mass) in the reaction plane is shown as a function of rapidity 
(scaled by the beam rapidity) for Au+Au reactions at impact parameters 
smaller than 4 fm (typical central collisions) and beam momentum 
of 4 GeV/c (upper window) and 12 GeV/c (lower window). Experiments at 
these energies have already been 
done, but no kaon flow analysis has been carried out yet. The filled 
circles are the results without any mean field potential for kaons. 
Comparing these to our early studies of nucleons and pions we see 
that kaons flow in the same direction as nucleons and pions. 
The magnitude of the average kaon transverse momentum in the reaction 
plane seems to be between that of nucleons and pions. More quantitatively, 
in the case of P/A=12 GeV/c, the average transverse momentum at the 
projectile rapidity is about 0.12,
0.04 and 0.01 GeV/c for nucleons, kaons and pions, respectively. 
With the kaon potential the flow pattern of kaons changes dramatically, 
especially around the midrapidity as one expects. This observation is 
in agreement with that found at lower energies\cite{gqli1,gqli2}, i.e., 
the kaon potential reduces the kaon flow or 
even changes its direction. This strong 
dependence on the kaon potential makes the kaon flow analysis a valuable 
tool to study the kaon dispersion relation in medium.   

To see more clearly the advantage of the kaon flow analysis over 
traditionally measured single particle observables in studying 
the kaon dispersion relation, we examine in Fig.\ 2 the kaon 
transverse momentum distribution around the midrapidity (left window) 
and projectile rapidity (right window) for the reaction of Au+Au at 
P/A=4 GeV/c and impact parameters less than 4 fm.
There seems to be no obviously measurable effect of the kaon potential 
on the inclusive spectra in both rapidity ranges. Similar results 
have also been seen at P/A=12 GeV/c and the calculations agree well 
with the E866/E802 data as we have shown in ref.\ \cite{art}.  
The observation that the kaon flow, but not the inclusive spectrum is 
sensitive to the kaon potential, is very similar to that observed for 
nucleons in heavy-ion collisions at Bevalac, SIS and AGS energies\cite{art}.  
The main reason for the insensitivity of the inclusive spectrum to the
mean-field potential is that the transverse collective 
flow velocity or the change of kaon momentum 
due to the mean-field potential, is too small compared to the average thermal 
velocity of particles to be seen in the inclusive spectrum.
For this reason it has been recently planned to measure the spectrum 
ratio of $K^+$ over $K^{-}$\cite{wolf}, where enhanced effects of $K^+$ and 
$K^-$ potentials, which are opposite in sign due to the G-parity, 
might show up more clearly\cite{fang}. 

In heavy-ion collisions at lower energies the azimuthal angle 
distribution ($dN/d\phi$) of particles
with respect to the reaction plane has been used to identify a 
possible collectivity perpendicular to the reaction plane.
The enhanced emission of particles in this direction has been 
dubbed ``squeeze-out'' phenomenon. For Au+Au reactions it was shown 
in ref.\cite{li94} that for both nucleons and pions the strength of
the squeeze-out decreases as the beam enegy increases and almost disappears
at energies higher than about 2 GeV/nucleon due to the 
increased global thermalization of the reaction system.
At AGS energies one thus expects kaons to have a very small 
collectivity perpendicular to the reaction plane.
Indeed, we find, as shown in Fig.\ 3 for the reaction of Au+Au at P/A=4 and 
12 GeV/nucleon, that there is no observable kaon squeeze-out 
from the calculations. Consequentlly, the azimuthal 
angle distribution with respect to the reaction plane is 
insensitive to the kaon potential. At lower energies, however, the
situation might be different\cite{gqli3}.  
 
In summary, within the framework of a relativistic transport model (ART) for 
heavy-ion collisions at AGS energies, we have examined the effects of kaon 
dispersion relation on the kaon transverse flow, transverse momentum 
distribution and azimuthal angle distribution.
It is found that the transverse flow is the most sensitive 
observable for studying the kaon dispersion relation in dense medium.

We would like to thank D. Best and G.Q. Li for useful discussions.
This work was supported in part by NSF Grant No. PHY-9509266.

\section*{Figure Captions}
\begin{description}
 
\item{\bf Fig.\ 1}\ \ \
The average transverse momentum of kaons in the reaction plane
for Au+Au reactions at $P_{beam}/A=$ 4 GeV/c (upper window) 
and 12 GeV/C (lower window) at impact parameters less than 4 fm. 
The open (filled) circles are the results 
obtained with (without) the kaon potential.

\item{\bf Fig.\ 2}\ \ \
Inclusive kaon transverse momentum distributions around midrapidity 
(left window) and projectile rapidity (right window) for the reaction of
Au+Au at $P_{beam}/A=$ 4 GeV/c and impact parameters less than 4 fm.
The open (filled) circles are the results 
obtained with (without) the kaon potential.

\item{\bf Fig.\ 3}\ \ \
The azimuthal angle distribution of kaons with respect to the reaction plane
in the Au+Au reactions described in Fig.\ 1. 
\end{description}
 
\end{document}